\documentclass[aps,prb,twocolumn,showpacs,preprintnumbers,amsmath,amssymb]{revtex4}

\usepackage[english]{babel}
\usepackage{graphicx}
\usepackage{subfigure}
\usepackage{dcolumn}
\usepackage{bm}
\usepackage{psfrag}

\begin{document}

\preprint{}

\title{Ab--initio low--energy dynamics of superfluid and solid $^4$He}
\author{E. Vitali}
 \affiliation{Dipartimento di Fisica, Universit\`a degli Studi
              di Milano, via Celoria 16, 20133 Milano, Italy}
\author{M. Rossi}
 \affiliation{Dipartimento di Fisica, Universit\`a degli Studi
              di Milano, via Celoria 16, 20133 Milano, Italy}              
\author{L. Reatto}
 \affiliation{Dipartimento di Fisica, Universit\`a degli Studi
              di Milano, via Celoria 16, 20133 Milano, Italy}
\author{D.E. Galli}
 \affiliation{Dipartimento di Fisica, Universit\`a degli Studi
              di Milano, via Celoria 16, 20133 Milano, Italy}
\date{\today}

\begin{abstract}
We have extracted information about real time dynamics of $^4$He systems
from noisy imaginary time correlation functions $f(\tau)$ computed via Quantum Monte Carlo (QMC):
production and falsification of model spectral functions $s(\omega)$ are obtained via a
survival--to--compatibility with $f(\tau)$ evolutionary process, based on Genetic Algorithms.
Statistical uncertainty in $f(\tau)$ is promoted to be an asset
via a sampling of equivalent $f(\tau)$ within the noise,
which give rise to independent evolutionary processes.
In the case of pure superfluid $^4$He we have recovered from exact QMC simulations
sharp quasi--particle excitations with spectral functions displaying also the multiphonon branch.
As further applications,
we have studied the impuriton branch of one $^3$He atom in liquid $^4$He and the
vacancy--wave excitations in hcp solid $^4$He finding a novel roton like feature.
\end{abstract}


\maketitle
\section{Introduction}
The development of {\it ab initio} theoretical descriptions 
of the low-energy dynamical behavior of quantum interacting models 
is naturally a very important issue in a huge variety of 
physical studies, ranging from Statistical Physics to Quantum Field Theory.
In the realm of Condensed Matter Physics, this requires
to start from the Hamiltonian operator $\hat{H}$ of a
many-body system and to investigate dynamical
properties via the study of {\it spectral functions}:
\begin{equation}
\label{spectral}
s(\omega) = \int_{-\infty}^{+\infty}\frac{dt}{2\pi}e^{i\omega t}\langle e^{i\hat{H} t}
\hat{A}e^{-i\hat{H} t}\hat{B}\rangle \quad ,
\end{equation}
$\hat{A}$ and $\hat{B}$ being given operators acting on the Hilbert space of the system,
and the brackets indicating expectation value on
the ground state or thermal average.
In this work we will address this topic in the case
of bulk $^4$He, which, during last decades,
has gained extreme interest since it provides
the simplest scenario in which quantum fluctuations 
and the statistics obeyed by the involved degrees of freedom govern
the physics of a macroscopic sample, giving rise to a big deal
of fascinating phenomena\cite{legg}.
The simple Hamiltonian of the system displays all the complexities
related to strong correlations among particles and has
been a very important test-ground both for many 
body theories and for numerical simulations.
In particular, the absence of the additional difficulties
connected with Fermi statistics has allowed Quantum Monte Carlo 
(QMC) methods to provide {\it exact} descriptions
of {\it equilibrium} phases of $^4$He, opening the possibility
of putting light into the intriguing physical mechanisms
underlying superfluidity and Bose Einstein condensation 
on a quantitative basis\cite{rmp}.

The natural idea of extending such approaches to {\it dynamical 
properties} (excitation spectra, transport coefficients etc.) is highly not trivial:
a {\it direct} QMC computation of \eqref{spectral} 
faces the problem of obtaining exact real time evolution,
and general solutions are not known.
Nevertheless, we can try to partially 
fill this {\it lack of knowledge} using
QMC techniques themselves. The stochastic processes related to imaginary
time Schroedinger equation underlying QMC simulations allow to perform
{\it observations} on the system, resembling
actual measurements on an experimental sample;
in particular in a QMC simulation it is
straightforward to collect a set of {\it observations}:
\begin{equation}
\label{valorimedi}
\mathcal{F} \equiv \{f_0,f_1,\dots,f_l\}
,\end{equation}
which are estimations of imaginary time correlation functions:
\begin{equation}
f(\tau) = \langle e^{\hat{H} \tau} \hat{A}e^{-\hat{H} \tau}\hat{B}\rangle
\end{equation}
in correspondence with a (unavoidably) finite number of
imaginary time values $\{0,\delta\tau,2\delta\tau,\dots,l\delta\tau\}$,
$\delta\tau$ being the time step of the QMC algorithm employed.
In general $\mathcal{F}$ is obtained as an average of several
QMC calculations of $f(\tau)$, each affected by statistical noise 
and which are used to estimate the
{\it statistical uncertainties} $\{\sigma_{f_0},\sigma_{f_1},\dots,\sigma_{f_l}\}$ 
associated with $\mathcal{F}$.

Such observations can provide information to {\it infer} an 
estimation of $s(\omega)$, through the exact relation:

\begin{equation}
\label{problem}
f(\tau) = \int_{-\infty}^{+\infty}d\omega\mathcal{K}(\tau,\omega)s(\omega)
\end{equation}
where for example, at zero temperature,
$\mathcal{K}(\tau,\omega) = \theta(\omega)e^{-\tau\omega}$, $\theta(\omega)$
being the Heaviside distribution. 
We have thus to face the {\it inverse problem}\cite{libromat}
of {\it deducing} the spectral function $s(\omega)$, inverting \eqref{problem}
starting from limited and noisy data.
At a first glance, one immediately convinces himself
that such an inverse procedure in most realistic situations is
unavoidably {\it ill--posed}, since any set of observations is
limited and noisy and the situation is even worse
since the kernel $\mathcal{K}(\tau,\omega)$
is a smoothing operator: the possibility of
finding out one and only one $s(\omega)$ {\it solving our problem}
is excluded.

Often sum rules provide useful help, either imposing exact constraints on
$s(\omega)$ or allowing to perform additional QMC measurements:
\begin{equation}
\label{moments}
\mathcal{C} \equiv \{\dots, c_0, c_1, \dots, c_n, \dots\} \quad ,
\end{equation}
which provide estimations for some moments of
$s(\omega)$:
\begin{equation}
\langle \omega^n \rangle = \int_{-\infty}^{+\infty} d\omega \omega^n s(\omega),\quad n\in\mathbb{Z}
.\end{equation}
For example $c_0$ is an estimation of $\langle \hat{A}\hat{B}\rangle$ which
may be easily obtained in equilibrium QMC simulations together with an associated statistical uncertainty.
Moreover some {\it a priori knowledge} may be assumed such as the support,
non--negativity or some further properties.

We would like to stress that the problem we have to face
belongs to the huge class of the {\it inverse problems}, which, 
since the earliest days of research in Physics, have always provided
challenges in a huge variety of physical or even more
generally scientific studies\cite{libromat,librophys}.
At the most general level, an inverse problem
emerges whenever one is building up a theoretical
description of a natural phenomenon and ought
to fill some {\it lack} of knowledge.
Typically one could need to infer some parameters of the theory,
and this could be achieved borrowing information from experimental data
or, as in our specific case, numerical {\it observations}, i.e. computer simulations.

The task of facing the problem in equation \eqref{problem}, typically referred in literature as an 
{\it analytic continuation} problem,
has already been investigated:
the Maximum Entropy Method\cite{mem1} (MEM) is the most widely popular strategy developed;
in the realm of bulk quantum fluids
it has provided only qualitatively interesting results\cite{mem2,rep}.
Other methods have been proposed: the Average Spectrum Method\cite{asm1} (ASM),
which has been recently applied
to lattice spin models\cite{asm2} but also to realistic off--lattice systems\cite{asm3},
the Stochastic Analytic Continuation (SAC) method\cite{sac1} and also the spectral analysis
described in Ref.\onlinecite{mpss}.
ASM and SAC are very similar approaches and it
has been proposed that MEM can be identified as a special limit of SAC\cite{sac2}.
Along this way, very recently a new algorithm strictly based on principles of Bayesian statistical
inference has been proposed\cite{sai}.
All these new approaches have been found able to reaveal some fine details of the spectral functions
but none of them has been applied to superfluid $^4$He which is the case study of this work.
We have used an inversion strategy which shares some features with the approaches cited above, but
possesses also some peculiar features: a novel way to deal with the statistical uncertainties
in the observations and the use of genetic algorithms to find spectral functions compatible with observations.
A preliminary application of this strategy to the determination of the
dynamical structure factor $S(q,\omega)$ of liquid $^4$He was presented in Ref.\onlinecite{procee};
here we explain it in detail and we present several applications to the Helium system.

The structure of the paper is the following. In section II we describe the
strategy which we have used;
in section III we show applications of this strategy on several Helium systems;
section IV contains our conclusions.
Appendix A contains some details of the used strategy, while in
Appendix B we present tests on the reliability of this strategy on known
spectral models.

\section{Inversion Strategy}
When considering ill-posed inverse problems
some important questions naturally arise: what can we really
learn when facing an inverse problem? What do we mean
when we speak about a {\it solution}?
In our opinion, a key proposition is brilliantly
put forward in Ref.\onlinecite{tarantola}, following Popper\cite{popper}:
{\it observations may be used only to falsify a theory}.
Translating this idea into the language of our problem,
we cannot expect to find out a recipe
that will allow to {\it deduce} all what is needed
to build up a unique function $s(\omega)$ from a given set of
observed {\it data}, maybe together with additional
informations, $\mathcal{D}=\left\{\mathcal{F},\mathcal{C}\right\}$. 
Nevertheless, provided that we are able to construct a
suitable parametrization of the abstract space $\mathcal{S}$
containing all the possible spectral functions,
we may use observations to provide a ``falsification test'', aimed to exclude
the functions $s(\omega)$ which fail to fit the data $\mathcal{D}=\left\{\mathcal{F},\mathcal{C}\right\}$
via \eqref{problem} and \eqref{moments}.
In this way we will be able to collect a (maybe very big) class of
spectral functions, which have been not falsified by the measured data.
In our opinion, the best way to achieve this is to fully
exploit all the information related to the observations:
that is, since any set of experimental data appears
together with statistical uncertainties evaluated
starting from suitable measurements, {\it any}
set of data compatible with the original one
has to take part to the falsification test, in
order to suppress the possibility of unphysical
effects arising from statistical fluctuations.

Once we remain with a set of equivalent spectral functions
``survived'' to the falsification test, depending on the mathematical
details of the space $\mathcal{S}$, a natural idea appears to be that of devising a procedure
allowing to capture what do the ``survived'' ones have in common.
In this way, even if we won't succeed in finding
out a unique $s \in \mathcal{S}$, we will be able nevertheless
to find out a class of features, providing physical properties, that $s$ has to possess
not to be falsified by the limited set of observations.
As explained below, to implement this
we need a space of models $\mathcal{S}$, containing a wide collection
of spectral functions consistent with any {\it prior knowledge} about $s(\omega)$,
a falsification procedure relying on the QMC ``measurements'' $\mathcal{D}=\{\mathcal{F},\mathcal{C}\}$ and
a strategy to capture the accessible physical properties of $s(\omega)$.
The strategy we are going to describe in the following relies on Genetic Algorithms\cite{goldberg}
to explore $\mathcal{S}$ and falsify its elements;
for this reason in the text we will refer to it as the Genetic Inversion via Falsification of Theories (GIFT)
strategy.

\subsection{The space of models}
In our mathematical framework $\mathcal{S}$ contains a wide class of step functions, 
providing a compromise between the possibility of suitably approximating
{\it any} model of spectral function and the feasibility of numerical operations
inside it.
In the typical case ($\hat{A}=\hat{B}^{\dagger}$) when $s(\omega)$ is known to be real-valued,
non-negative and the zero-moment sum-rule holds, we rely on models $\overline{s}$ of the form:
\begin{equation}
\label{solutions2}
\overline{s}(\omega) = \sum_{j=0}^{N_\omega-1}\frac{s_j}{\mathcal{M}\Delta\omega}\chi_{A_j}(\omega),
\quad \sum_{j=0}^{N_\omega-1}s_j = \mathcal{M} \quad .
\end{equation}
$\overline{s}(\omega)$ differs from
the physical spectral functions by a factor $c_0$,
the zero--moment sum rule,
which belongs to the set of observations and its role will become evident below.
We introduce a discretization of the codomain:

\begin{equation}
\label{discretization}
s_j \in \mathbb{N}\,\cup\{0\}
, \end{equation}
to make the space finite, and we use the characteristic function $\chi_{A_j}(\omega)$  
of the intervals $A_j = [\omega_j,\omega_{j+1})$, 
$\{\omega_0,\dots,\omega_{N_\omega}\}$
being a partition of width $\Delta\omega$ of an interval of the real line
larger than the hypothesized support of $s(\omega)$.
$\mathcal{M}$ provides the maximum number of quanta of spectral weight available
for the ensemble of the intervals $A_j$.

\subsection{The falsification principle}
Once we have defined the space of model spectral functions, we have to devise
a practical strategy to implement the falsification principle.
We have to rely on the QMC estimations $\mathcal{D}=\{\mathcal{F},\mathcal{C}\}$.
To keep the description simpler we now concentrate only on $\mathcal{F}$;
naturally all what we will say refers also to $\mathcal{C}$ with obvious modifications.
The numbers $\{f_0,f_1,\dots,f_l\}$
are averages evaluated during a simulation and appear together with their 
estimated statistical uncertainties $\{\sigma_{f_0},\sigma_{f_1},\dots,\sigma_{f_l}\}$.
In typical approaches, such information are dealt with inside the framework of Bayes' theorem;
they provide the key ingredients to build up the {\it a posteriori} probability\cite{mem1} to be
maximized, together with some {\it a priori} probability, to extract the {\it most probable}
spectral function.

On the other hand, we find natural to suggest a novel way of exploiting the
information contained in $\{\sigma_{f_0},\sigma_{f_1},\dots,\sigma_{f_l}\}$:
any set $\mathcal{F}^{\star}$ {\it equivalent} to \eqref{valorimedi},
i.e. such that  $|f^{\star}_i - f_i|$ is of the same order as $\sigma_{f_i}$,
could be a result of another simulation. 
Falsifying the elements of $\mathcal{S}$ should require 
not only compatibility with $\mathcal{F}$ but also compatibility with a vast population
of $\mathcal{F}^{\star}$ equivalent to the set \eqref{valorimedi} of data.
In general, relying on independent simulations to generate {\it equivalent} sets $\mathcal{F}^{\star}$
could represent a very demanding computational task; thus,
when this is not sistematically practicable, we need a recipe to generate {\it equivalent} sets $\mathcal{F}^{\star}$,
and then we have to use the generated $\mathcal{F}^{\star}$ to falsify
the elements of $\mathcal{S}$.
At the simplest level we have addressed the generation of the sets $\mathcal{F}^{\star}$ by 
sampling independent Gaussian distributions centered on the original observations,
with variances corresponding to the estimated statistical uncertainties. 
A generic element $\mathcal{F}^{\star}$ is then:
\begin{equation}
\label{star}
\mathcal{F}^{\star} \equiv \{f_0 + \varepsilon^{\star}_0, f_1 + \varepsilon^{\star}_1, \dots,
f_l + \varepsilon^{\star}_l \}= \{f^{\star}_0, f^{\star}_1, \dots, f^{\star}_l \}
\end{equation}
being $\varepsilon^{\star}_j$ random numbers sampled from Gaussian distributions
with zero mean and variances equal to $\sigma^2_{f_j}$.
See the end of Appendix A to read about possible extensions related to this point.
When the procedure in \eqref{star} has been used, a posteriori, in some selected cases,
one can check the accuracy of the results obtained
by comparing these with models coming from the analysis of independent QMC observations.
We stress that the very idea of exploiting the statistical uncertainties in the 
observations for generating {\it equivalent} sets $\mathcal{F}^{\star}$ is
the main difference with respect to preexisting statistical approaches to inverse problems.

The key point is then to falsify the elements of $\mathcal{S}$ relying on each one of
these sets\eqref{star}: compatibility means small deviations from the observations. Thus,
given the set $\mathcal{F}^{\star}$, a very simple measure of the compatibility of a
model with this set of observations can be obtained by computing
\begin{equation}
\Delta(\overline{s}) = \frac{1}{l+1}
\sum_{j=0}^l \left[ f_j^{\star} - \int d\omega \, e^{-\omega j \delta\tau}
c_0^{\star}\,\overline{s}(\omega)\right]^2 \quad.
\label{deviation}
\end{equation}
The normalization of our models requires the multiplication of $\overline{s}(\omega)$ by
the estimation, $c_0$, of the zero moment, which belongs to the set of observations $\mathcal{D}$;
consistently, we sample also its value analogously to how we treat $\mathcal{F}$.
This is the reason why a factor $c_0^{\star}$ appears in \eqref{deviation}.

Each member $\mathcal{F}^{\star}$ of equivalent data leads to a different model;
let us call $\overline{s}_k$ the model found with the $k$-th member $\mathcal{F}^{\star}$.
Each one of these models cannot be trusted to be
the solution of the inverse problem,
being at least partially biased by the particular $\mathcal{F}^{\star}$;
in other terms we can say that each one of these models will posses
spurious information, presumably different in each model, together with some physical information.
An averaging procedure is therefore the simplest way to filter out the spurious information
and to reveal physical information, which consist in the common features among the
models which have not been falsified:
\begin{equation}
S_{GIFT}(\omega) = \frac{1}{\mathcal{N}_{r}} \sum_{k=1}^{\mathcal{N}_{r}} c_0^{(k)}\overline{s}_k(\omega)
\label{reconstructed}
\end{equation}
where $\mathcal{N}_{r}$ is the number of equivalent random set of $\mathcal{F}^{\star}$ used in the computation
and $c_0^{(k)}$ is the $c_0^{\star}$ used in the $k$-th reconstruction.
We stress again that the averaging procedure in \eqref{reconstructed} does not represent the absence
of a sensible strategy for the choice among the generated $\overline{s}_k$; 
in fact, as explained in
Ref.\onlinecite{tarantola}, the whole collection of
the not falsified models should be considered.
Contrarily, we have a sensible strategy: we must not make any choice,
all the models that have not been falsified are equivalent; as a consequence we are
interested only in their common features, an information that we
extract via the averaging procedure in \eqref{reconstructed}.

The average procedure in the definition of $S_{GIFT}(\omega)$
points toward some similarities between our strategy and that of ASM or SAC and also that
of Ref.\onlinecite{mpss}.
However, in the light of the falsification principle, these approaches are fairly different:
in order to obtain their ``solution'', ASM and SAC average over spectral functions obtained
by exploring model-space regions via a local Metropolis random walk based on a probability
distribution;\cite{asm1,sac1}
in these approaches the statistical uncertainties in the observations play a different role,
appearing only in the definition of the probability (ASM and SAC) or in the definition of the 
{\it minimal deviation} in Ref.\onlinecite{mpss}.
Another issue is the algorithm used to explore the space of models; as explained below, GIFT uses
a {\it non-local} dynamics induced by a stochastic evolutionary process instead of a local
Metropolis random walk which, in principle, could suffer from ergodicity problems,
being the high probability model-space regions not guaranteed to be connected.

At this point the following question urges an answer:
How can we practically explore $\mathcal{S}$?
We have implemented genetic algorithms as efficient algorithms to explore our huge space
of models, $\mathcal{S}$. There could be inverse problems and different space of models which
could be more efficiently explored with other algorithms; obviously, the 
``inversion via falsification of theories'' approach, which consists mainly
in the novel treatment of the statistical uncertainties of observations, can be applied 
also using a ``dynamics'' in the space of models different from the genetic one.

\subsection{The fitness and the genetic dynamics}
Genetic algorithms (GA) provide an extremely efficient tool 
to explore a sample space by a {\it non-local} stochastic dynamics, 
via a survival--to--fitness evolutionary process
mimicking the natural selection we observe in natural world; such evolution
aims toward ``good'' {\it building blocks}\cite{goldberg} which,
in our case, should recover information on physical spectral functions.
The fitness of one particular $\overline{s}(\omega)$ should be based
on the observations, i.e., on the noisy measured set $\mathcal{D}=\{\mathcal{F},\mathcal{C}\}$.
But as explained before, taking into account the estimated statistical noise of $\mathcal{D}$,
any set $\mathcal{D}^{\star}$ compatible with $\mathcal{D}$ provides equivalent information
to build a fitness function.
Thus in our GA any random set $\mathcal{D}^{\star}=\{\mathcal{F}^{\star},\mathcal{C}^{\star}\}$\cite{footnote}
gives rise to a fitness, which simply compares
``predictions'' of theories and ``observations'':
\begin{equation}
\Phi_{\mathcal{D}^{\star}}(\overline{s}) =
-\Delta(\overline{s})
- \sum_n \gamma_n \left[c_n^{\star} - \int d\omega \,\omega^n
\,c_0^{\star}\,\overline{s}(\omega)\right]^2
\label{fitness}
\end{equation}
In \eqref{fitness} the free parameters $\gamma_n > 0$ are adjusted in order to make the
contributions to $\Phi_{\mathcal{D}^{\star}}$ coming from $\mathcal{F}^{\star}$ and from
$\mathcal{C}^{\star}$ of the same order of magnitude:
the idea is that we are not allowed to prefer some particular observation among the others,
thus they should have the same weight in the fitness.
If it happens that one moment is exactly known, no error is added making $c_n^{\star}=\langle \omega^n \rangle$.
We stress that \eqref{fitness} provides the simplest and the most natural definition;
moreover, as explained below, our GA uses $\Phi_{\mathcal{D}^{\star}}$ only to order
models in ascending fitness, thus any alternative definition of $\Phi_{\mathcal{D}^{\star}}$
which provides the same ordering will give rise to an identical genetic dynamics.

GA are well know procedures characterized by well defined (genetic--like) operations
on populations of candidate solution to optimization problems in applied mathematics.
For basic nomenclature and 
standard implementations one can refer to textbooks (e.g. see Ref.\onlinecite{goldberg}).
In Appendix A we present our particular realization related to the space of models
we have defined.
In our GA, we start randomly constructing a collection of $\overline{s}(\omega)$;
each $\overline{s}(\omega)$ is coded by $N_{\omega}$ integers, $s_j$ in equation \eqref{solutions2}.
The genetic dynamics then consists in a succession
of {\it generations} during which an initial {\it population},
consisting of $\mathcal{N}_{\overline{s}}$ {\it individuals}, is
replaced with new ones in order to reach regions of $\mathcal{S}$
where high values of the {\it fitness} exist, for a given $\mathcal{D}^{\star}$.
In practice, in the passage between two generations a succession 
of ``biological--like'' processes takes place, and namely
{\it selection}, {\it crossover} and {\it mutation}.
The {\it selection} procedure is meant to choose preferentially
individuals with large fitness in the process of producing
the next generation (see Appendix A for more details).
 
In our context the GA dynamics performs the falsification procedure:
only the $\overline{s}(\omega)$ with the highest fitness in the last generation provides
a model, $\overline{s}_{k}(\omega)$, which has not been falsified by $\mathcal{D}^{\star}$.
The maximum amount of generations, $\mathcal{N}_{\mathcal{G}}$, is chosen in order to
reach the condition $\Delta(\overline{s}(\omega))\simeq \delta$ (See Appendix A).
Many independent evolutionary processes are generated by sampling different
$\mathcal{D}^{\star}$, thus obtaining a set made of the
elements $c_0^{(k)}\,\overline{s}_{k}(\omega)$;
at this point, as explained above, the averaging procedure \eqref{reconstructed} 
extracts the common
features in this set and this produces the GIFT estimate of the spectral function.

\section{Results for Helium systems}
We are ready now to present applications of our approach
on physical systems.
Long Monte Carlo runs have been performed in order to get imaginary time correlation functions
with a typical statistical uncertainty of 0.1-1\%.
For bulk superfluid $^4$He
most of the simulations have been performed with $N = 64$ and $N = 256$ atoms moving in a cubic box,
but also $N = 128$ and $N = 512$ have been studied;
for solid $^4$He the hcp lattice with $N = 180$ and $N = 448$ lattice positions have been used.
Imaginary time correlation functions have been computed
for instants $\tau_l = l\delta\tau$, $l = 0,.., l_{max}=60$ in the superfluid phase
and $l_{max}=30$ in the solid phase, spaced by $\delta\tau = 1/160$ K$^{-1}$.
All the results shown in this article have been obtained with the interatomic
interaction of Ref.\onlinecite{aziz},
but some computations have been performed also with that in Ref.\onlinecite{aziz95}
as mentioned in the text.
We have used the pair--product approximation\cite{rmp} to express the
imaginary time propagator in the interval $\delta\tau = 1/160$ K$^{-1}$ which is known to be very accurate.
For bulk superfluid $^4$He we choose $\gamma_n = 0$ $\forall n \neq 1$ (see equation \eqref{fitness}),
i.e., we have included only $c_0$, which is the estimation of the static structure factor,
and the first moment sum--rule which is exactly known, $\langle \omega \rangle =|\vec{q}|^2/2m$.
For the extraction of the impurity branch and of the vacancy excitation spectrum we have only
used the zero moment sum rule.
Other parameters were fixed to $\Delta \omega=0.25$ K, $\mathcal{M}=5000$, $N_\omega=600-1600$,
initial value of $\mathcal{N}_{\overline{s}}=25000$
which is decreased down to the minimum value
$\mathcal{N}_{\overline{s}}=400$, as explained in Appendix A;
we have used about $10^3$ different sets $\mathcal{D}^{\star}$
and the number of generations for a given $\mathcal{D}^{\star}$
have been fixed to $10^4$.
We have performed many tests with different choices of
such parameters showing that none has a critical role under condition that
$N_\omega\Delta\omega$ is larger of the support of the reconstructed spectral functions.
\subsection{The dynamical structure factor of superfluid $^4$He}
\begin{figure}[t]
 \includegraphics[width=8cm]{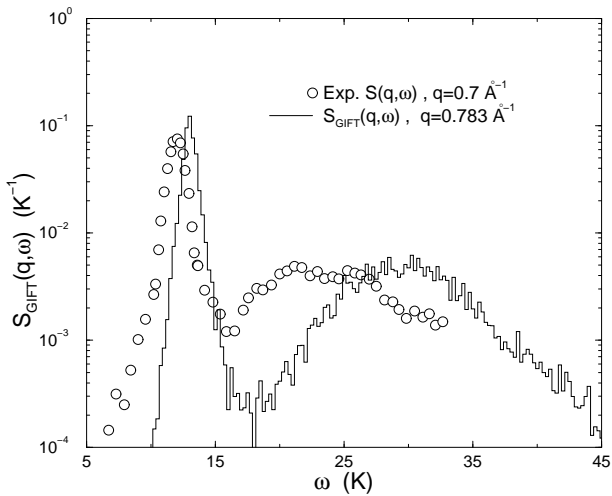}
 \caption{\label{fig1}
(line) $S_{GIFT}(q,\omega)$ for $q=0.783$ \AA$^{-1}$ and $\rho=0.0218$ \AA$^{-3}$;
(open circles) observed\cite{andersen} dynamic structure factor $S(q,\omega)$ in liquid $^4$He
for $q=0.7$ \AA$^{-1}$ at SVP and $T=1.3$ K. Notice the logarithmic scale.
Notice also the difference between the wave vector of $S_{GIFT}(q,\omega)$ and the one of the 
experimental available\cite{andersen} dynamic structure factor; the experimental single particle
peak position is known to increase by about $0.8$ K in moving from $q=0.7$ \AA$^{-1}$ to 
$q=0.783$ \AA$^{-1}$.
}
\end{figure}

Our first case study is the determination of the
dynamical structure factor $S(q,\omega)$ of liquid bulk $^4$He.
We have used the exact SPIGS method\cite{spigs1,spigs2}
to compute the intermediate scattering function $F(q,\tau)$
at $T = 0$ K near the equilibrium density, $\rho=0.0218$ \AA$^{-3}$,
and slightly above the freezing density, $\rho=0.0262$ \AA$^{-3}$; 
$F(q,\tau)$ is simply $f(\tau)$
when $\hat{A}=\hat{B}^{\dagger}$ is chosen to be the Fourier transform of the local
density operator $\hat{A} = \hat{\rho}_{\vec{q}} = \sum_{i=1}^{N}e^{-i\vec{q}\cdot \vec{\hat{r}}_i}$.
An example of our reconstructed $S_{GIFT}(q,\omega)$ is shown in Fig.\ref{fig1},
it exhibits an overall structure
in good agreement with experimental data: a sharp quasi--particle
peak and a shallow multiphonon maximum are present.
Both features appear for the first time within an analytic continuation procedure applied to a QMC study
of superfluid $^4$He.
Notice that it is not appropriate to compare the widths of the two sharp quasi--particle peaks
in Fig.\ref{fig1}: in fact the experimental peak includes the broadening arising from instrumental
resolution and the effect of the finite temperature; on the contrary, as explained in the following,
the width of the reconstructed GIFT peak from a $T=0$ 
imaginary time correlation function is mainly a measure
of the uncertainty in reconstructing its position.
\begin{figure}[t]
 \includegraphics[width=8cm]{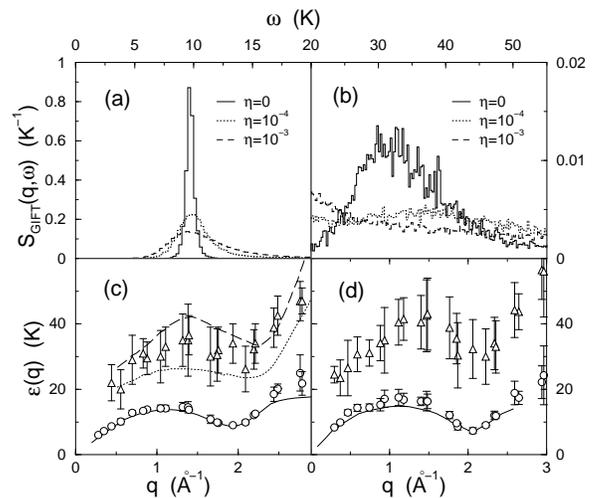}
 \caption{\label{fig2} 
(a) and (b): $S_{GIFT}(q,\omega)$ at $q=1.755$ \AA$^{-1}$ and $\rho=0.0218$ \AA$^{-3}$;
(a) single quasi--particle (qp) peak; (b) multiphonon (mp) contribution (notice change of scale).
Lines corresponding to a $S_{GIFT}(q,\omega)$ obtained with a nonzero entropic prior 
($\eta\neq 0$) are also shown.
(c) $\varepsilon(q)$ extracted at $\rho=0.0218$ \AA$^{-3}$
from the position of the qp (circles)
peaks and the positions of the maxima of the mp contribution (triangles) are shown.
The error--bars represent the 1/2--height widths. 
(d) $\varepsilon(q)$ and mp contribution extracted at $\rho=0.0262$ \AA$^{-3}$.
Lines in (c) and (d): experimental data\cite{cowley,cowley2};
in the mp region in (c) the lower curve (dotted)
represents the position of the maximum while the upper one (dashed)
represents the 1/2--height width.
}
\end{figure}
In Fig.\ref{fig2} we show one $S_{GIFT}(q,\omega)$ in the roton region
together with the excitation energies $\varepsilon(q)$ i.e., the position of the main peak as
function of $q$. The uncertainties of $\varepsilon(q)$
correspond to the widths of the peaks $\sigma_\varepsilon$:
we have checked the consistency of such identification by performing
independent QMC estimations of $F(q,\tau)$ and comparing the positions of the peaks
obtained in $S_{GIFT}(q,\omega)$; the distribution of the peaks displays a variance
comparable to $\sigma_\varepsilon^2$.

In principle also a MEM-like algorithm could fit into the GIFT approach:
it is enough to modify the fitness function
by adding to $\Phi_{\mathcal{D}^{\star}}$ in equation \eqref{fitness}
an entropic term $-\eta S$, with
\begin{equation}
S = \int d\omega \left\{ \overline{s}(\omega) \ln\left[ \frac{\overline{s}(\omega)}{m(\omega)} \right] 
- \overline{s}(\omega) + m(\omega)
\right\} \quad ,
\end{equation}
$S$ being the entropy as in Ref.\onlinecite{mem2} and $\eta \ge 0$ a free parameter;
$m(\omega)$ is the default model which in previous implementations\cite{mem2,rep}
has been chosen to be simply a constant in absence of any prior knowledge.
This is not a faithful implementation of MEM because the entropic term
is used in the context of GIFT and not within the framework
of Bayes' theorem.
Anyway, it provides results for the dynamical structure factor of superfluid
$^4$He very similar to those appeared in literature\cite{mem2,rep}:
by using a constant as default model, $m(\omega)$,
for all wave vectors $\vec{q}$ we observed for the main peak of $S(q,\omega)$ a
broadening (see Fig.\ref{fig2})
strongly dependent on the choice of the parameter $\eta$.
This makes us loose a great deal of information and makes the extracted excitation energies
critically dependent on the value of $\eta$, thus introducing ambiguities
in the interpretation of the results.
Recently, a new fully Bayesian approach has been proposed\cite{sai}, which avoids
ad--hoc assumptions on the relative intensity of the entropic term and which is able to
reconstruct spectral functions with more pronounced features.
It will be interesting in the future to see how this new method or other recent
Bayesian methods perform on superfluid $^4$He.
Given their ability in reconstructing some fine details of the spectral functions,
observed in studying different quantum systems, it is possible that 
such methods will give equivalent or even
better results than GIFT when applied to the same inverse problem.
In our original approach, i.e. without $\eta S(\overline{s})$,
we have checked that none of the parameters
(like $\mathcal{M}$, $\Delta\omega$, $\alpha$, $\gamma_n$, ...) affects
the class of features that we may trust to carry reliable
physical information.

\begin{figure}[t]
 \includegraphics[width=8cm]{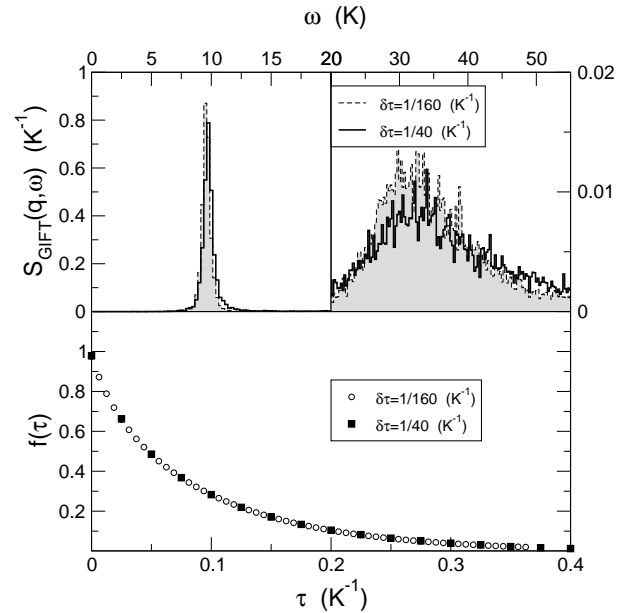}
 \caption{\label{fig2b}
Upper panels: $S_{GIFT}(q,\omega)$ at $q=1.755$ \AA$^{-1}$ and $\rho=0.0218$ \AA$^{-3}$
extracted from noisy imaginary time correlation functions with different level of accuracy;
(left) single quasi--particle peak; (right) multiphonon contribution (notice change of scale).
Lower panel: imaginary time correlation functions $f({\tau})$ used in the GIFT reconstructions shown in
the upper panel.
}
\end{figure}
In Fig.\ref{fig2b} (see the upper panel) we compare the spectral function shown in the upper panels of
Fig.\ref{fig2} with a spectral function extracted with GIFT from a more noisy
correlation function (see lower panel in Fig.\ref{fig2b}) computed with a less accurate imaginary time propagator
for instants $\tau_l = l\delta\tau$, $l = 0,.., l_{max}=17$, spaced by $\delta\tau = 1/40$ K$^{-1}$.
In this new GIFT reconstruction the statistical uncertainties $\{\sigma_{f_0},\sigma_{f_1},\dots,\sigma_{f_l}\}$
are about four times bigger, i.e. about $4\times 10^{-3}$ instead of about $10^{-3}$, but even if we
have less accurate observations on about four times fewer imaginary time points, GIFT is able to reconstruct
a spectral function displaying an elementary excitation peak and a multiphonon contribution
in agreement with the result of the more accurate simulation.
This shows the robustness of GIFT against less accurate QMC data.
Further studies on the robustness of GIFT against inaccurate QMC data
are shown in Appendix B, where tests on
known spectral models are presented (see Fig.\ref{graf5}).

\begin{figure}[t]
 \includegraphics[width=8cm]{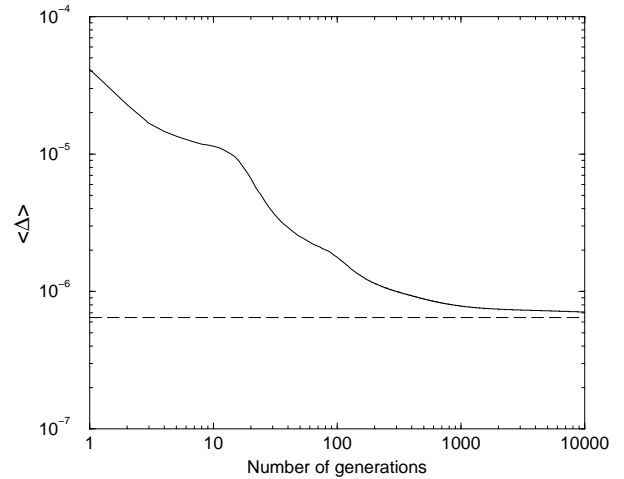}
 \caption{\label{grafnew}
Evolution of the deviation \eqref{deviation} during the stochastic evolution
of the genetic algorithm for the reconstruction plotted in Fig.\ref{fig2}(a,b) for $\eta=0$
averaged  with respect to the sampled sets $\mathcal{D}^{\star}$.
The dashed horizontal line represents the value $\delta=\frac{1}{l+1}\sum_{j=0}^l \sigma^2_{f_j}$.
}
\end{figure}
As an example of the stochastic evolution of a GIFT computation,
in Fig.\ref{grafnew} we show the deviation \eqref{deviation} as a function
of the number of generations in the evolutionary process for the reconstruction
plotted in Fig.\ref{fig2}(a,b) for $\eta=0$ averaged on the sampled sets $\mathcal{D}^{\star}$.
One can see that the maximum number of generations, $\mathcal{N}_{\mathcal{G}}$,
we have used in this reconstruction is optimal in reaching
the ``compatibility'' condition, $\Delta(\overline{s})\simeq\delta=\frac{1}{l+1}\sum_{j=0}^l \sigma^2_{f_j}$,
without overfitting (this point is expanded in Appendix A).

By integrating $S_{GIFT}(q,\omega)$ with respect to $\omega$  in the range of the sharp peak
and in the remaining frequancy range we have access to the
strength of the single quasi--particle peak, $Z(q)$, and to the contribution to
the static structure factor, $S(q)$, coming from multiphonon excitations.
\begin{figure}[t]
 \includegraphics[width=8cm]{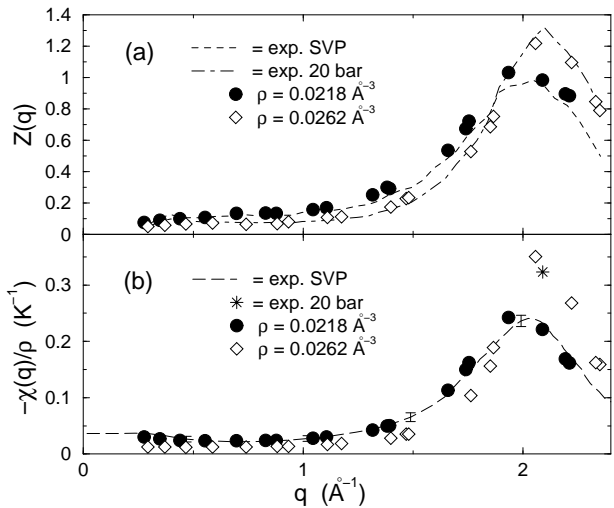}
 \caption{\label{fig3}
(a) GIFT strength of the quasi particle peak $Z(q)$ as function of $q$ at two densities and
experimental data\cite{gibbs}.
(b) GIFT Static density response function $\chi (q)$ at two densities and
experimental data\cite{cowley,caupin}.
Error bars of theoretical results are smaller than the symbol size.}
\end{figure}
Remarkably, $Z(q)$ turns out to be in close
agreement with experimental data (see upper Fig.\ref{fig3}), thus
strongly suggesting that the shallow maximum in $S_{GIFT}(q,\omega)$ at large energy
carries indeed reliable physical information on the multiphonon
branch of the spectrum. The position of such multiphonon maximum (see Fig.\ref{fig2}c) is in
qualitative agreement with experiments\cite{cowley}:
as we show in Appendix B, 
within the present implementation of GIFT
there is no possibility to recover the detailed
shape of the spectral function like the multiphonon substructures given by 
high resolution measurements\cite{gibbs} of $S(q,\omega)$.
In the lower panel of Fig.\ref{fig3} we show the static density response function $\chi (q)$
obtained evaluating the $\langle \omega^{-1} \rangle$ from $S_{GIFT}(q,\omega)$; the agreement with experiments
is impressive, also near freezing\cite{caupin}. 

The calculation of the excitation spectrum $\varepsilon(q)$ in superfluid $^4$He
via QMC was addressed for instance in Ref.\onlinecite{moro} and in Ref.\onlinecite{boro}, but
here we are clearly much more ambitious because we aim to reconstruct
the full spectral function.
In our reconstructed spectral functions the elementary excitation peaks are so accurately
resolved that it is possible
to reveal the effects of even fine details of the interatomic interaction.
For example, the computed spectrum $\varepsilon(q)$ in the phonon region is about $0.7$ K above the
experimental value. We understand this as an effect of truncation of the inter--atomic
interaction $v(r)$ at a certain distance $r_c$. In most of our computations
the interatomic potential is cut-off and displaced to zero at
$r_c = 6$ \AA, and the equation of state gives rise to an overestimation of
the sound velocity by about 16\%.
We have performed some computations with $r_c = 14$ \AA, in a simulation of $N=512$
$^4$He atoms and in this case the sound velocity turns out to be correct and now
the theoretical $\varepsilon (q)$ at small $q$ agrees with
experiment within the resolution $\Delta \omega$.

In order to give a more detailed description of the roton region we have computed
$\varepsilon (q)$ for many wave vectors in the roton region
and the average of the excitation energies nearby the roton minimum,
produces our estimate of the roton energy, $E_R$, as shown in Table \ref{rotons}.
\begin{table}
\begin{center}
\begin{tabular}{ccccccc}

$\rho$ (\AA$^{-3}$) && $E^I_R$ (K) &&  $E^{II}_R$ (K) && Experimental $E_R$ (K) \\


\hline          

0.0218  && $8.96 \pm 0.47$ && $8.67 \pm 0.29$ && $8.608 \pm 0.01$ \\
0.0262  && $7.43 \pm 0.34$ && $7.22 \pm 0.27$ && $7.3 \pm 0.02$ \\


\hline     
\end{tabular}
\end{center}
\caption{\label{rotons} Roton energies, $E_R$, at two different densities and using the
$v(r)$ in Ref.\onlinecite{aziz}, $E^I_R$, and the $v(r)$ in Ref.\onlinecite{aziz95},
$E^{II}_R$, a potential considered more accurate.
In the last column experimental data\cite{stir} are shown.
}
\end{table}

\subsection{Impurity and vacancy dynamics}
Another interesting test case is provided by liquid $^4$He
in presence of one $^3$He impurity, in order to extract
the impurity branch which has been
experimentally measured\cite{fak}. 
Variational results for such branch are known\cite{gisa} but no results from exact QMC
are available.
This calculation requires the choice
of $\hat{A} = e^{-i\vec{q}\cdot \vec{r}_{imp}}$, where $\vec{r}_{imp}$
is the position of the impurity.
\begin{figure}[t]
 \includegraphics[width=8cm]{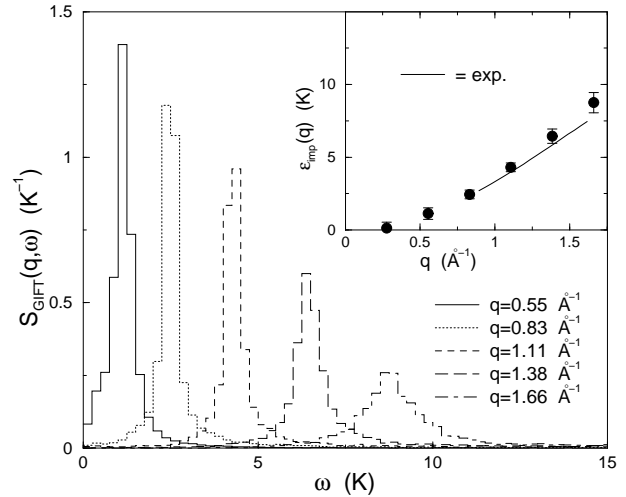}
 \caption{\label{fig4}
Impurity $^3$He quasi-particle peak in superfluid $^4$He at SVP for several wave vectors;
in the inset the extracted excitation energies
are shown together with experimental data\cite{fak}.
}
\end{figure}
In Fig.\ref{fig4} we show the reconstructed spectral functions
together with the estimated dispersion relation obtained
from a simulation of $N = 255$ $^4$He atoms and one $^3$He atom at $\rho=0.0218$ \AA$^{-3}$.
The agreement with experimental data \cite{fak} is very good, thus
providing a robust check of validity of our approach.

As a further application of GIFT we have studied
the excitation spectrum of a single vacancy in hcp solid $^4$He at $\rho=0.0293$ \AA$^{-3}$,
a density slightly above melting.
The behavior of vacancies in solid $^4$He is of high interest because vacancies 
and other defects are believed to have a key role in the possible supersolid phase
of $^4$He at low temperature\cite{chan,chan2}.
In order to apply GIFT to vacancy dynamics the first step is the definition of a
vector position $\vec{x}_v$ that allows to follow
the ``motion'' of the vacancy in imaginary time during a SPIGS simulation.
This problem is much more difficult than
the evaluation of the impurity branch, because the very definition
of $\vec{x}_v$ is far from trivial due to the large zero--point
motion of the atoms in the low density solid.
$\vec{x}_v$ turns out to be a many--body variable,
depending on all the vector positions of $^4$He atoms, and
even not free of ambiguities.
We have employed two different procedures to obtain $\vec{x}_v$:
the coarse-grain\cite{prokov} and the Hungarian\cite{hung,cepev}.
\begin{figure}[t]
 \includegraphics[width=8cm]{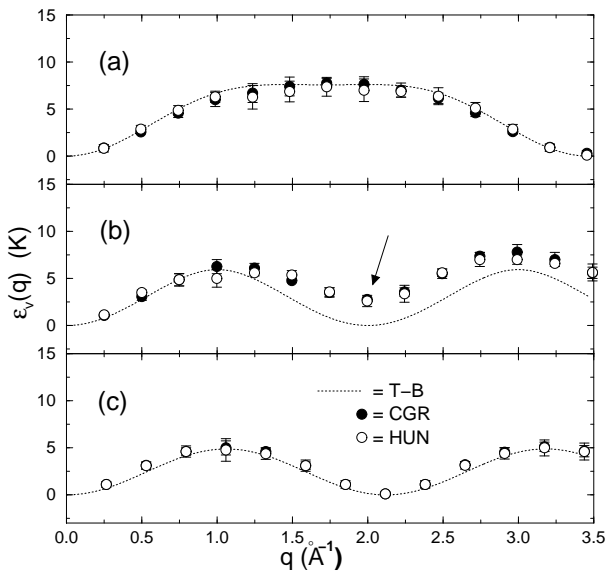}
 \caption{\label{fig5}
Vacancy excitation spectrum in solid $^4$He extracted from the $S_{GIFT}(q,\omega)$
of the vacancy--vector position $\vec{x}_v$ at
$\rho=0.0293$ \AA$^{-3}$ in a hcp lattice with N=447 particles along the
principal symmetry directions: (a) $\Gamma$K, (b) $\Gamma$M and (c) $\Gamma$A.
Two different algorithms have been used to obtain $\vec{x}_v$:
a coarse-grain algorithm\cite{prokov} (CGR)
and the Hungarian algorithm\cite{hung,cepev} (HUN).
Dotted lines represent the spectrum of a tight binding model (T--B)
for the hcp lattice\cite{gallisv}
obtained imposing the values of the band width along the $\Gamma$K and $\Gamma$A directions
equal to the average between the values extracted from the two different algorithms.
The arrow points out the vacancy--roton mode.
}
\end{figure}
In Fig.\ref{fig5} we show the vacancy excitation spectrum $\varepsilon_v(\vec{q})$
extracted from the vacancy spectral functions ($\hat{A}=e^{-i\vec{q}\cdot\vec{x}_v}$)
obtained with the two methods.
The results obtained with the two definitions of $\vec{x}_v$
are very similar, and at first sight make evident
a picture of Bloch waves in the crystal; the agreement with a tight binding
hopping model\cite{gallisv} is good.
Notice that $\varepsilon_v(\vec{q})$ represents the excitation energy with respect to
the state with a vacancy with $|\vec{q}|=0$, i.e., $\varepsilon_v(\vec{q})$
does not include the vacancy activation energy.
By fitting $\varepsilon_v(\vec{q})$ with the tight binding expression we extract the 
vacancy effective mass in the different lattice directions:
$m^\star_{\Gamma K}=m^\star_{\Gamma M}=0.46 \pm 0.03 m_4$ and $m^\star_{\Gamma A}=0.55 \pm 0.1 m_4$,
where $m_4$ is the $^4$He mass; these values for $m^\star$
are in agreement with the results obtained with a different method in Ref.\onlinecite{prokovac}.

The agreement of $\varepsilon_v(\vec{q})$ with the tight binding model fails dramatically
in the $\Gamma$M direction.
In fact, at any reciprocal lattice vector the excitation energy should vanish
and this agrees with our results along the $\Gamma$K and $\Gamma$A directions as one can see in Fig.\ref{fig5}.
On the contrary at the first reciprocal lattice vector along
$\Gamma$M our vacancy excitation spectrum does 
not vanish but it reveals a novel vacancy--roton mode with an energy of $2.6 \pm 0.4$ K
and an effective mass of about $m_R^{\star}=0.46 \, m_4$.
We have checked that this energy does not depend on the size of the system.
Such behavior of $\varepsilon_v(\vec{q})$ in the $\Gamma$M direction implies
that the (non--zero) minimum is not a consequence of the lattice periodicity
but it is related to correlated motion of particles like in superfluid $^4$He.
It is interesting that neutron scattering from hcp $^4$He gives an unexpected
excitation mode beyond the phonon modes exactly in the $\Gamma M$ direction with
a roton--like mode at the reciprocal wave vector\cite{goodk}.
The experimental energy of such roton mode is about 4.4 times larger than what we find;
so it is unclear the connection, if any, between our mode and experimental data.
A larger vacancy roton energy might arise in presence of clusters of vacancies.
By analyzing the contributions to 
$f(\tau) = \langle e^{\hat{H} \tau} \hat{A}e^{-\hat{H} \tau}\hat{A}^{\dagger}\rangle$
with $\hat{A}=e^{-i\vec{q}\cdot\vec{x}_v}$, 
one can see that the vacancy--roton mode is connected to motions of the vacancy between different 
basal planes. The fundamental difference between in--basal--plane and inter--basal--plane motions
is that the lattice position in the first case is a centre of inversion whereas this
is not so in the second case. The fact that hcp is not a Bravais lattice is fundamental
in this respect.
We have verified that in bcc crystal and in a two dimensional triangular lattice,
both Bravais lattices, no such vacancy roton mode is present.

\section{Conclusions}
We have extracted information about the dynamics of a quantum many--body systems via
analytic continuation of QMC data, obtaining
very accurate results in the $^4$He case, in the liquid and in the solid
phase, even in presence of disorder. Our results provide major improvements
with respect to previous MEM studies appeared in literature on superfluid $^4$He:
we have been able to recover sharp quasi--particle excitations, with excitation energies
in good agreement with experimental data, and spectral functions
displaying also the multiphonon branch with the correct relative spectral weight.
As discussed in the Introduction, the ability to reveal some fine details of the spectral
functions has been already observed in more recent Bayesian methods applied to other systems.
These methods have never been applied to 
the $^4$He case; it will be interesting in the future to compare the results of
all these different strategies applied to the same inverse problem.

The basic idea of the falsification principle\cite{tarantola} guided us to follow a particular strategy
which relies on Genetic Algorithms to explore the space of models
to find those of them which are compatible with observations.
Each of these models is affected by the noise and by the limited information on the dynamics
of the system but we identify the relevant physical information by extracting
the features that are common to such compatible models.
This is obtained via an averaging procedure among the collection of models which has not been falsified.
This feature of the
strategy we have used has some similarities with other methods\cite{asm1,asm2,sac1, mpss}
but significant differences are present in the role of statistical noise.
A drawback of this strategy is the repeated
computationally demanding exploration of the space of models via the genetic dynamics;
this can be faced much more efficiently by implementing simultaneous falsification procedures on different
sets of observations $\mathcal{D}^\star$ via parallel computation.
Anyway, it remains the possibility that different inverse problems and/or different space of models
could be more efficiently explored with algorithms different from Genetic Algorithms.
The used analytic continuation strategy can be extended to include
different kinds of constraints on the spectral function
or additional information like cross correlations between the statistical noise of
$f(\tau)$ at different imaginary times;
many variants of it can be devised depending on the problem, for instance a basis set different from
step functions \eqref{solutions2} can be used or non uniform discretization
in presence of problems with multiple time scales, or distribution of noise that is not Gaussian.

\section{Acknowledgments}
We acknowledge useful discussions with S. Moroni, A. Motta and M. Nava.
This work has been supported by Regione Lombardia and CILEA Consortium
through a LISA Initiative (Laboratory for Interdisciplinary Advanced Simulation) 2010 grant
[link: http://lisa.cilea.it].

\appendix
\section{Details and possible extensions of GIFT}
The implemented {\it selection} procedure in our GA choose preferentially
individuals with large fitness by
ordering the population in ascending fitness
and {\it selecting} the $k$-th individual with
\begin{equation}
\label{selection}
k=\left[\mathcal{N}_{\overline{s}}\,\,r^{1/3}\right]+1
\end{equation}
where $r$ is an uniform random number, $r \in [0,1)$, and
$\left[\dots\right]$ is the integer part;
the non linear dependence of $k$ on $r$ ensures
that individuals with large fitness are preferentially selected.
The {\it crossover} then operates on two selected $\overline{s}(\omega)$,
the {\it father} and the {\it mother},
exchanging subparts of their total number of quanta of spectral weight, $\mathcal{M}$,
to generate two {\it sons}.
We have used a special single point crossover by sampling a random integer, $w$,
between 0 and $\mathcal{M}$ and by exchanging $w$ randomly chosen quanta of spectral weight
between the {\it father} and the {\it mother}.
In this way, the second equation in \eqref{solutions2} is automatically satisfied, implying
that the zero--moment sum rule is also satisfied.
Each exchanged quantum remains in the original frequency bin as in its parents, thus
ensuring that strong features present in both parents tend to persist in the sons.
Successively, with a given probability, {\it  mutation} takes place
on the two new individuals, i.e., a shift
of a fraction of spectral weight between two intervals $A_j$.
This is repeated till a new generation of $\overline{s}(\omega)$ replaces the old one,
with the exception of the $\overline{s}(\omega)$ with the highest fitness in the old generation which
is cloned (elitism).
The number of individuals in the new population is constantly reduced by about 5\% at every generation
till $\mathcal{N}_{\overline{s}}$ is equal to a given minimal value; from this point over, the number
of individuals $\mathcal{N}_{\overline{s}}$
in the new generations is kept constant to this minimum value.
The discarded individuals are those with the smallest fitness in the population.
This is done to start the genetic evolution from a wide variety of possible models
without dissipating computational time on falsified spectral functions.

The choice of the form of $\Delta(\overline{s})$ in \eqref{deviation} is not critical because,
as explained above, its role in GIFT is only to assign
an ordering among models based on their compatibility with observations;
thus alternative definitions of $\Delta(\overline{s})$ that do not
change this ordering will give rise to identical results.
However, in presence of strong variations among the estimated statistical
uncertainties $\{\sigma_{f_0},\sigma_{f_1},\dots,\sigma_{f_l}\}$ it is preferible in the definition of
$\Delta(\overline{s})$ to divide each term of the sum by the relative $\sigma_{f_j}^2$
in order to give more weight to more precise observations.
The statistical uncertainties of the imaginary time correlation functions, computed
in our studies of $^4$He systems, were found quantitatively comparable;
therefore, we have used $\Delta(\overline{s})$ as defined in \eqref{deviation}.

The natural scale of
$\Delta(\overline{s})$ is provided by the value $\delta=\frac{1}{l+1}\sum_{j=0}^l \sigma^2_{f_j}$:
models $\overline{s}$ such that $\Delta(\overline{s}) << \delta$ may provide unphysical
overfitting.
In our statistical approach to inverse problems there are two procedures which preserve
from overfitting. The first one is that, given a set $\mathcal{F}^{\star}$,
the exploration of the space of models $\mathcal{S}$
should be stopped when a model $\overline{s}(\omega)$ is found such that $\Delta(\overline{s}) \simeq \delta$;
a further reduction of $\Delta$ will only represents the intention to give to
$\mathcal{F}^{\star}$ a strong belief, which is incompatible with the statistical treatment
of the observations in our strategy.
The second procedure is even more relevant and in some sense it is intrinsic to our strategy:
given an $\mathcal{F}^{\star}$ the reconstructed
model $\overline{s}(\omega)$ contains some
spurious information, but these information will be averaged out in $S_{GIFT}(\omega)$.

In the present applications on $^4$He systems
the whole covariance matrix has not been computed, thus equivalent sets $\mathcal{F}^\star$
have been sampled simply by using the procedure in equation \eqref{star}.
In general, the knowledge of the whole covariance matrix should not be neglected;
however, in Appendix B we show (see Fig.\ref{graf6}) that even the exact
knowledge of the correlation function on an equivalent or slightly wider discrete
set of imaginary time instants, $\tau_l$,
is not enough to substantially improve the reconstruction abilities of our strategy
when the kernel in equation \eqref{problem} is of the form
$\mathcal{K}(\tau,\omega) = \theta(\omega)e^{-\tau\omega}$.
Thus in the present study we have considered only the diagonal part
of the covariance matrix in order to reduce the computational cost of our QMC simulations
and GIFT reconstructions. 
We have also checked that this limitation does not seem to affect the present results by performing
a few GIFT reconstructions using outputs of several independent simulations, which
represent the exact (but computationally heavy) sampling of $p(\mathcal{F}^\star)$, instead of constructing
$\mathcal{F}^\star$ via equation \eqref{star}.
Note that in presence of more complete information in the observations, like
an estimation of the full covariance matrix $\Sigma$,
for the data in \eqref{valorimedi},
the generation of the equivalent sets $\mathcal{F}^{\star}$
can be readily generalized by sampling an $(l+1)$-variate normal distribution
with the following probability density function:
\begin{equation}
\label{multivariate}
p(\mathcal{F}^{\star})=\frac{
\exp \left[ -\frac{1}{2} (\mathcal{F}^{\star}-\mathcal{F})^\intercal \Sigma^{-1}
(\mathcal{F}^{\star}-\mathcal{F}) \right]}
{(2\pi)^\frac{l+1}{2}\det(\Sigma)^\frac{1}{2}}
\quad ;
\end{equation}
standard methods to perform efficiently this task are known (see for example Ref.\onlinecite{gentle}).

In the present applications we have not explored
different variants of GIFT as, for instance, a basis set different from
step functions; one cannot exclude the possibility that by using different variants
more information could be obtained.

\section{Tests on known spectral models}
Here we show several tests of application of GIFT
on known analytical spectral models suitably discretized
and ``dirtied'' with random noise to ``simulate'' actual data.
It will appear evident what we have already pointed out
in the introduction: only some features of the exact solution
can be consistently reproduced;
we have no possibility to reconstruct exactly the shape of $s(\omega)$;
on the other hand, access is granted to the
identification of the presence of peaks and to their
positions, to some integral properties involving $s(\omega)$ and to its support.
\begin{figure}[t]
 \includegraphics[width=8cm]{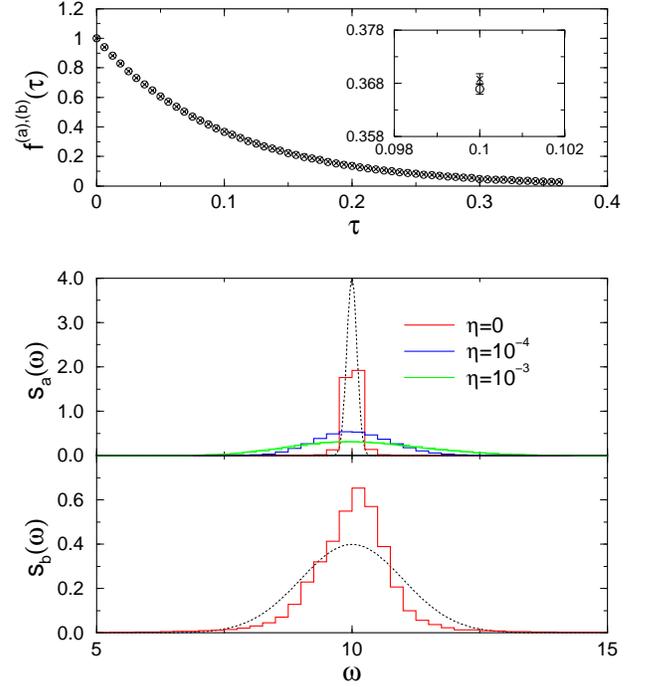}
 \caption{\label{graf1}
(Color online) Single peak reconstruction.
Upper panel: Two noisy imaginary time correlation functions obtained via \eqref{data} from $f^{(a)}(\tau)$
(open circles) and $f^{(b)}(\tau)$ (x symbols), which are the
Laplace transforms of $s_a(\omega)$ (dotted line in the middle panel: $\mu=10$ and $\alpha=0.1$)
and $s_b(\omega)$ (dotted line in the lower panel: $\mu=10$ and $\alpha=1$).
The inset is a zoom on one imaginary time instant.
Middle panel: $s_a(\omega)$ (dotted line) and reconstructed $S_{GIFT}(\omega)$ (red line)
using in the fitness $\Phi_{\mathcal{D}^{\star}}$ only the first moment (i.e. $\gamma_n=0$
$\forall n\neq 1$); green and blue lines represent MEM-like reconstructions with different
values of the $\eta$ parameter in the fitness (see legend).
Lower panel: $s_b(\omega)$ (dotted line) and reconstructed $S_{GIFT}(\omega)$
using in the fitness $\Phi_{\mathcal{D}^{\star}}$ only the first moment (i.e. $\gamma_n=0$
$\forall n\neq 1$).
}
\end{figure}

The most natural test for the reliability of the GIFT approach
is provided by a systematic study of Laplace
inversion problems whose analytical solution is known.
Our idea is to focus our attention on model functions of the form:
\begin{equation}
\label{test:model}
s(\omega) =\theta (\omega)\sum_{j=1}^{\mathcal{N}_p}
p_j\frac{e^{-\frac{(\omega - \mu_j)^2}{2\alpha_j^2}}}{\sqrt
{2\pi}\alpha_j} 
\quad \sum_{j=1}^{\mathcal{N}_p} p_j = 1
\end{equation}
linear combinations of Gaussians multiplied by $\theta (\omega)$,
the Heaviside distribution, resembling qualitatively the experimental
results for spectral functions in condensed matter physics at $T=0$. 
We may perform several tests varying the parameters $\mathcal{N}_p$,
number of maxima, $\{\mu_1, \dots, \mu_{\mathcal{N}_p}\}$, positions of the
maxima, $\{\alpha_1, \dots, \alpha_{\mathcal{N}_p}\}$, widths of the
peaks and $\{p_1, \dots, p_{\mathcal{N}_p}\}$, the areas under the peaks.
The Laplace transform $f(\tau)$ of \eqref{test:model} may be expressed in terms of the
standard complementary error function:
\begin{equation}
\operatorname{erfc}(z) = \frac{2}{\sqrt{\pi}}
\int_{z}^{+\infty}dt e^{-t^2} \quad ,
\end{equation}
whose values are tabulated, in the following form:
\begin{equation}
\label{laptest}
f(\tau) =\frac{1}{2}
\sum_{j=1}^{\mathcal{N}_p} p_je^{-\mu_j \tau + \frac{\tau^2\alpha_j^2}{2}}
\operatorname{erfc}\left(\frac{\tau\alpha_j - \mu_j/\alpha_j}{\sqrt{2}}\right)
\end{equation}
In order to {\it simulate} the output of a typical QMC calculation, we define the 
{\it measured imaginary time data} $\mathcal{F} = 
\{f_0,\dots,f_l\}$ as:
\begin{equation}
\label{data}
f_j = f(j\delta\tau) + \varepsilon_j
\end{equation}
where $f(j\delta\tau)$ is evaluated from \eqref{laptest}, and 
$\varepsilon_j$ are random numbers, mimicking the error bars
affecting QMC data, following Gaussian distributions
with zero mean and variances, $\sigma^2_{\varepsilon_j}$, comparable with 
the ones typically occurring in our QMC results ($\sigma_{\varepsilon_j}/f_j$ in the range 0.1--4 \%). 
$f_j$ play the role of the output of QMC simulation; GIFT falsification uses $\mathcal{N}_{r}$ random sets
$\mathcal{F}^{\star}=\{f^{\star}_0,\dots,f^{\star}_l\}$ defined by
\begin{equation}
\label{data2}
f^{\star}_j = f_j + \varepsilon^{\star}_j
\end{equation}
$\varepsilon^{\star}_j$ being Gaussian random variables with zero mean and variances which here,
to be coherent with the applications we have presented,
we assume to be equal to $\sigma^2_{\varepsilon_j}$.
\begin{figure}[t]
 \includegraphics[width=8cm]{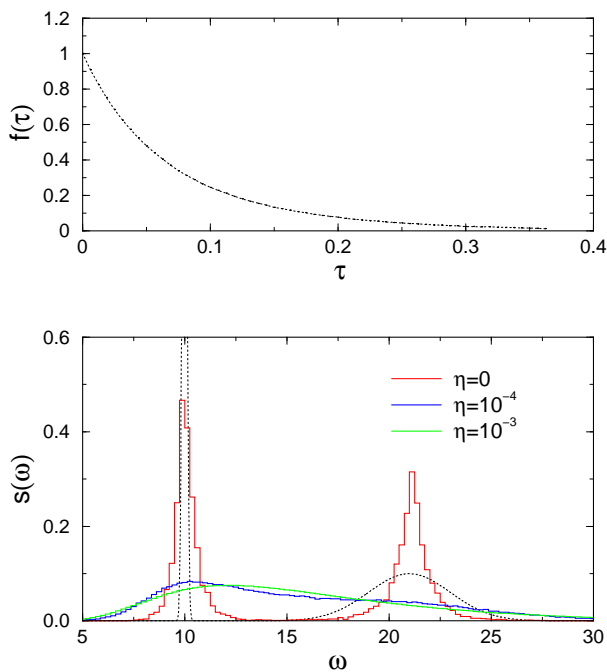}
 \caption{\label{graf2}
(Color online) Double peak reconstruction for well separated peaks.
Upper panel: Noisy imaginary time correlation function obtained via \eqref{data} from $f(\tau)$
(dotted line) which is the Laplace transform of $s(\omega)$ (dotted line in the lower panel,
see text for parameters).
Lower panel: $s(\omega)$ (dotted line) and reconstructed $S_{GIFT}(\omega)$ (red line) from $f(\tau)$
using in the fitness $\Phi_{\mathcal{D}^{\star}}$ only the first moment (i.e. $\gamma_n=0$
$\forall n\neq 1$); green and blue lines represent MEM-like reconstructions with different
values of the $\eta$ parameter in the fitness (see legend).
}
\end{figure}

Our aim is to compare \eqref{test:model} with the GIFT result we obtain
pretending that our knowledge about the imaginary time correlation function is limited to the 
discretized and noisy data 
$\mathcal{F}$ in \eqref{data}, and
to other available information $c_n$ about the moments
which, inside these tests on analytically solvable models, can be evaluated from
\eqref{test:model}; we will neglect now the error bars affecting the values of
the $c_n$.
The parameters we have employed in our GIFT reconstructions
are listed in Table \ref{tabpar}.
Obviously, the choices of the interval of the frequency space, of the resolution $\Delta\omega$
(which fixes $N_{\omega}$), of the discretization $\Delta\tau$ and of the number
of points in imaginary time $l$ are crucial for a specific spectral function one is trying
to reconstruct and should be chosen consistently with the considered model;
the other parameters are not crucial for a correct functioning of GIFT
and they have been chosen in order to falsify a wide variety of models leaving
the computational cost of the algorithm at a reasonable level.
\begin{figure}[t]
 \includegraphics[width=8cm]{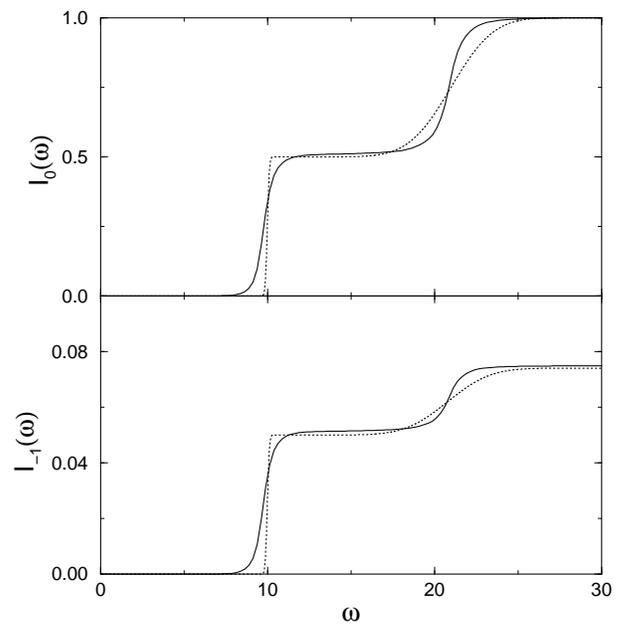}
 \caption{\label{graf2b}
Integral properties for double peak reconstruction.
Upper panel: $I_0(\omega)$ from the exact $s(\omega)$ (dotted line) and $I_0(\omega)$
obtained with the reconstructed $S_{GIFT}(\omega)$ in Fig. \ref{graf2} (case $\eta=0$).
Lower panel: $I_{-1}(\omega)$ from the exact $s(\omega)$ (dotted line) and
$I_{-1}(\omega)$ obtained with the reconstructed $S_{GIFT}(\omega)$ in Fig. \ref{graf2}.
}
\end{figure}

\begin{table}
\begin{center}
\begin{tabular}{cccc}

&name of the parameter &  symbol & value\\


\hline          

SM&number of bins in frequency space  &  $N_{\omega}$    & $600$ \\
SM&resolution in frequency space      &  $\Delta\omega$  & $0.25$ \\
SM&number of quanta of spectral weight&  $\mathcal{M}$   & $5 \times 10^3$ \\
CF&discretization in imaginary time   &  $\delta\tau$    & $1/160$ \\ 
CF&number of points in imaginary time &  $l$      & $60$ \\
GA&number of generations              &  $\mathcal{N}_{\mathcal{G}}$ & $10^4$ \\
GA&initial number of models      &  $\mathcal{N}_{\overline{s}}$ & $2.5 \times 10^4$ \\
GA&final number of models       &  $\mathcal{N}_{\overline{s}}$   & $400$ \\ 
GA&number of new random sets generated&  $\mathcal{N}_{r}$        & $10^3$     \\


\hline     
\end{tabular}
\end{center}
\caption{\label{tabpar} Typical parameters used with GIFT related to: the space of models (SM),
the correlation function (CF) and the genetic algorithm (GA).
}
\end{table} 

Also in the reconstruction of
known models of spectral functions
one can compare GIFT results with those based on
the strategy of MEM by adding in the fitness function
an entropic term, as we did with
the dynamical structure factors in superfluid $^4$He from QMC imaginary time
correlation functions.

\subsection{Single peak reconstruction}
The simplest test--case is provided by the attempt of reconstructing spectral
functions displaying only one peak at a given point $\mu$ with a width $\alpha$.
The upper panel of Fig. \ref{graf1} makes evident the difficulty of the
inverse problem: two functions with the same
parameter $\mu_a = \mu_b = 10$ but different values of the widths, 
respectively $\alpha_a = 0.1$ and $\alpha_b = 1.0$, in imaginary
time domain differ by about $0.5$\%, of the same order as the typical QMC 
error bars. It is clear then that the information about the width of the peak
is always strongly obscured by the noise. However, from the middle and
lower panel of Fig. \ref{graf1} it is manifest that GIFT reconstruction,
obtained using in the fitness $\Phi_{\mathcal{D}^{\star}}$ only the first moment (i.e. $\gamma_n=0$
$\forall n\neq 1$), is able to capture with an high accuracy the position of the peak 
in both cases. We note that, despite the difficulty mentioned above, in this
simple case of a single peak, even the widths are remarkably semi--quantitatively recovered.  
This ability is evidently lost when the entropic term with a constant default model is 
switched on (see Fig.\ref{graf1}) with values of $\eta$ similar to those used in
Fig.\ref{fig2}.

\subsection{Double peak reconstruction}
In order to get closer to realistic physical applications, we try to
reconstruct also spectral functions displaying a double peak.
Inside such a double peak reconstruction, we may check also the 
estimation of the integrated spectral functions:
\begin{eqnarray}
\label{integral}
I_0(\omega) = \int_{0}^{\omega}d\omega's(\omega') \quad , \nonumber \\
I_{-1}(\omega) = \int_{0}^{\omega}d\omega'\frac{s(\omega')}{\omega'} \quad .
\end{eqnarray}
$I_0(\omega)$ provides information about the spectral weight under the peaks
in $s(\omega)$; in particular in the $\omega$ range between the two peaks
$I_0(\omega)$ gives the information from which we have derived the strength
of the single quasi--particle peak, $Z(q)$, in our GIFT study of superfluid $^4$He,
as we will show in the following section.
On the other hand, the asymptotic value of $I_{-1}(\omega)$ for large $\omega$
provides the key to estimate the static response function $\chi(q)$.
\begin{figure}[t]
 \includegraphics[width=8cm]{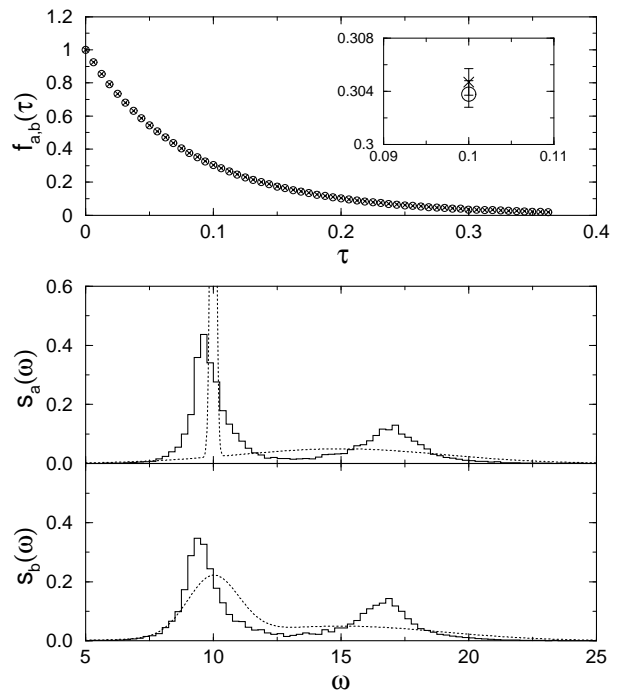}
 \caption{\label{graf3}
Double peak reconstruction for overlapping peaks.
Upper panel: Two noisy imaginary time correlation functions obtained via \eqref{data} from $f^{(a)}(\tau)$
(open circles) and $f^{(b)}(\tau)$ (x symbols), which are the
Laplace transforms of $s_a(\omega)$ (dotted line in the middle panel) and $s_b(\omega)$
(dotted line in the lower panel); see text for parameters.
The inset is a zoom on one imaginary time instant.
Middle panel: $s_a(\omega)$ (dotted line) and reconstructed $S_{GIFT}(\omega)$
using in the fitness $\Phi_{\mathcal{D}^{\star}}$ only the first moment (i.e. $\gamma_n=0$
$\forall n\neq 1$).
Lower panel: $s_b(\omega)$ (dotted line) and reconstructed $S_{GIFT}(\omega)$
using in the fitness $\Phi_{\mathcal{D}^{\star}}$ only the first moment (i.e. $\gamma_n=0$
$\forall n\neq 1$).
}
\end{figure}

In Fig. \ref{graf2} we show a reconstruction of a spectral function $s(\omega)$
for two well separated peaks
($p_1=0.5$, $p_2=0.5$, $\mu_1=10$, $\mu_2=21$, $\alpha_1=0.1$ and $\alpha_2=2.0$)
using in the fitness $\Phi_{\mathcal{D}^{\star}}$ only the first moment (i.e. $\gamma_n=0$
$\forall n\neq 1$); this is the typical fitness used in our reconstruction of spectral functions
of superfluid $^4$He.
The corresponding $I_{0}(\omega)$ and $I_{-1}(\omega)$
are plotted in Fig. \ref{graf2b}
compared with the analytic results from \eqref{test:model}.
We observe that no appreciable difference emerges, with respect to the exact results, as far as the
determination of the positions of the peaks, of the
areas under the peaks, and of the $\langle \omega^{-1} \rangle$ 
moment (see Fig. \ref{graf2b}) are concerned: the accuracy
is very good; on the other hand,
the shape of the reconstructed $s(\omega)$ has not to be taken too seriously because
it belongs to the class of properties whose determination is obscured
by statistical errors and discretization in imaginary time.
For values of $\eta$ similar to those used in Fig.\ref{fig2},
MEM-like reconstructions are not even able to detect the presence of two peaks; moreover
the position of the maximum of the reconstructed spectral function is dangerously $\eta$--dependent
(see Fig. \ref{graf2}) thus showing the importance to find a strategy which avoids
ad--hoc assumptions\cite{sai}.

In Fig. \ref{graf3} we consider two different spectral functions $s_a(\omega)$ and $s_b(\omega)$,
characterized by two overlapping peaks,
whose Laplace transforms, in imaginary time domain, are plotted in the upper panel
($p_{1a}=0.5$, $p_{2a}=0.5$, $\mu_{1a}=10$, $\mu_{2a}=15$, $\alpha_{1a}=0.1$ and $\alpha_{2a}=4.0$;
 $p_{1b}=0.5$, $p_{2b}=0.5$, $\mu_{1b}=10$, $\mu_{2b}=15$, $\alpha_{1b}=1.0$ and $\alpha_{2b}=4.0$)
As discussed previously, the small difference, comparable with the (pretended) error bars,
rules out the possibility of a reconstruction of the actual shapes. Nevertheless,
GIFT succeeds in finding out the positions of the peaks with good
accuracy even in this case in which the overlap between the two peaks becomes significant.

\subsection{Multiple peak reconstruction}
\begin{figure}[t]
 \includegraphics[width=8cm]{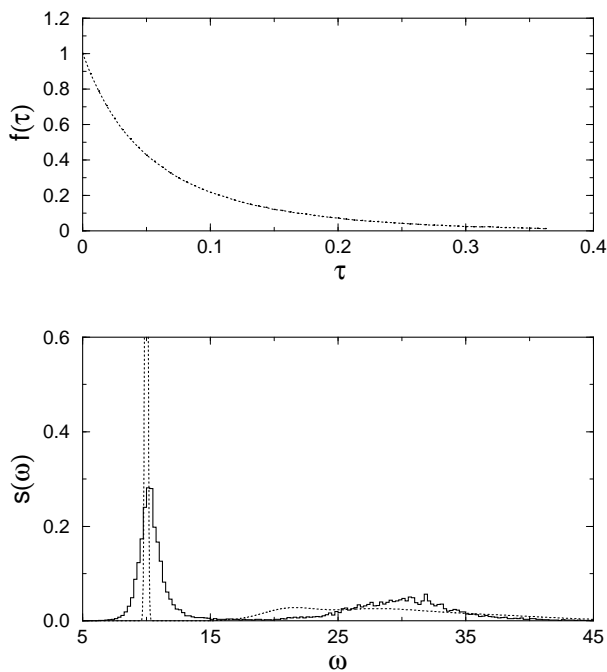}
 \caption{\label{graf4}
Multiple peak reconstruction.
Upper panel: Noisy imaginary time correlation function obtained via \eqref{data} from $f(\tau)$
which is the Laplace transform of $s(\omega)$ (dotted line in the lower panel).
Lower panel: $s(\omega)$ (dotted line) and reconstructed $S_{GIFT}(\omega)$ from $f(\tau)$
using in the fitness $\Phi_{\mathcal{D}^{\star}}$ only the first moment (i.e. $\gamma_n=0$
$\forall n\neq 1$).
}
\end{figure}
\begin{figure}
 \includegraphics[width=8cm]{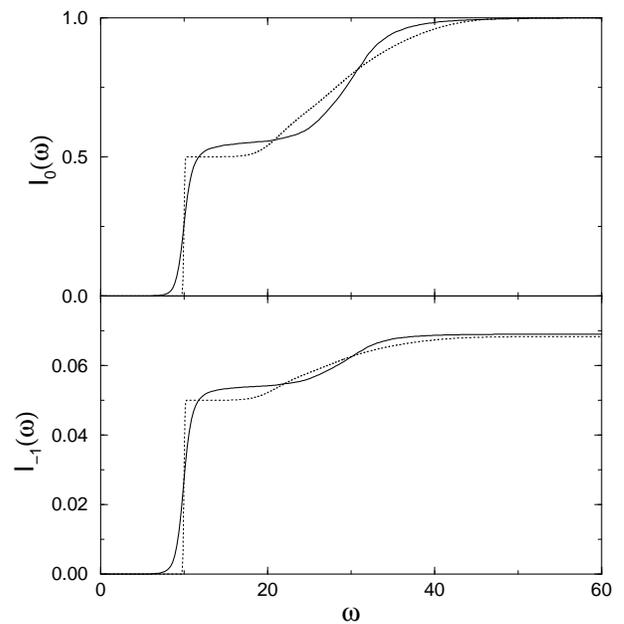}
 \caption{\label{graf4b}
Integral properties for multiple peak reconstruction.
Upper panel: $I_0(\omega)$ from the exact (dotted line) and $I_0(\omega)$ obtained with the
reconstructed $S_{GIFT}(\omega)$ in Fig. \ref{graf4}.
Lower panel: $I_{-1}(\omega)$ from the exact (dotted line) and $I_{-1}(\omega)$ obtained with the
reconstructed $S_{GIFT}(\omega)$ in Fig. \ref{graf4}.
}
\end{figure}
\begin{figure}
 \includegraphics[width=8cm]{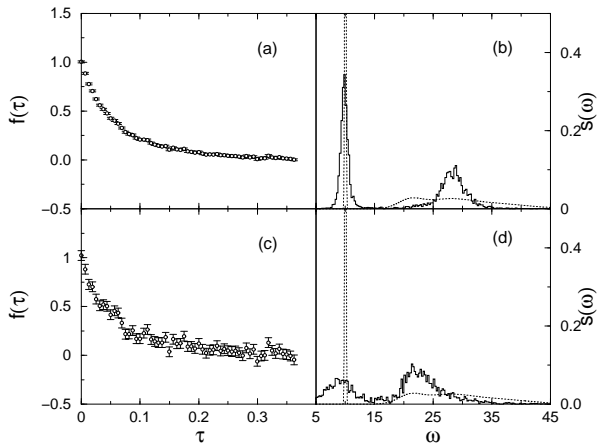}
 \caption{\label{graf5}
Panel (a): $f_j$ values obtained with $\sigma_{\varepsilon_j}=0.01$ and used by GIFT
to reconstruct $s(\omega)$ as shown in panel (b).
Panel (c): $f_j$ values obtained with $\sigma_{\varepsilon_j}=0.05$ and used by GIFT
to reconstruct $s(\omega)$ as shown in panel (d).
Panel (b): exact $s(\omega)$ (dotted line) as in Fig. \ref{graf4} and reconstructed
$S_{GIFT}(\omega)$ using the noisy observation of $f(\tau)$ in panel (a).
Panel (d): exact $s(\omega)$ (dotted line) as in Fig. \ref{graf4} and reconstructed
$S_{GIFT}(\omega)$ using the noisy observation of $f(\tau)$ in panel (c).
}
\end{figure}
\begin{figure}
 \includegraphics[width=8cm]{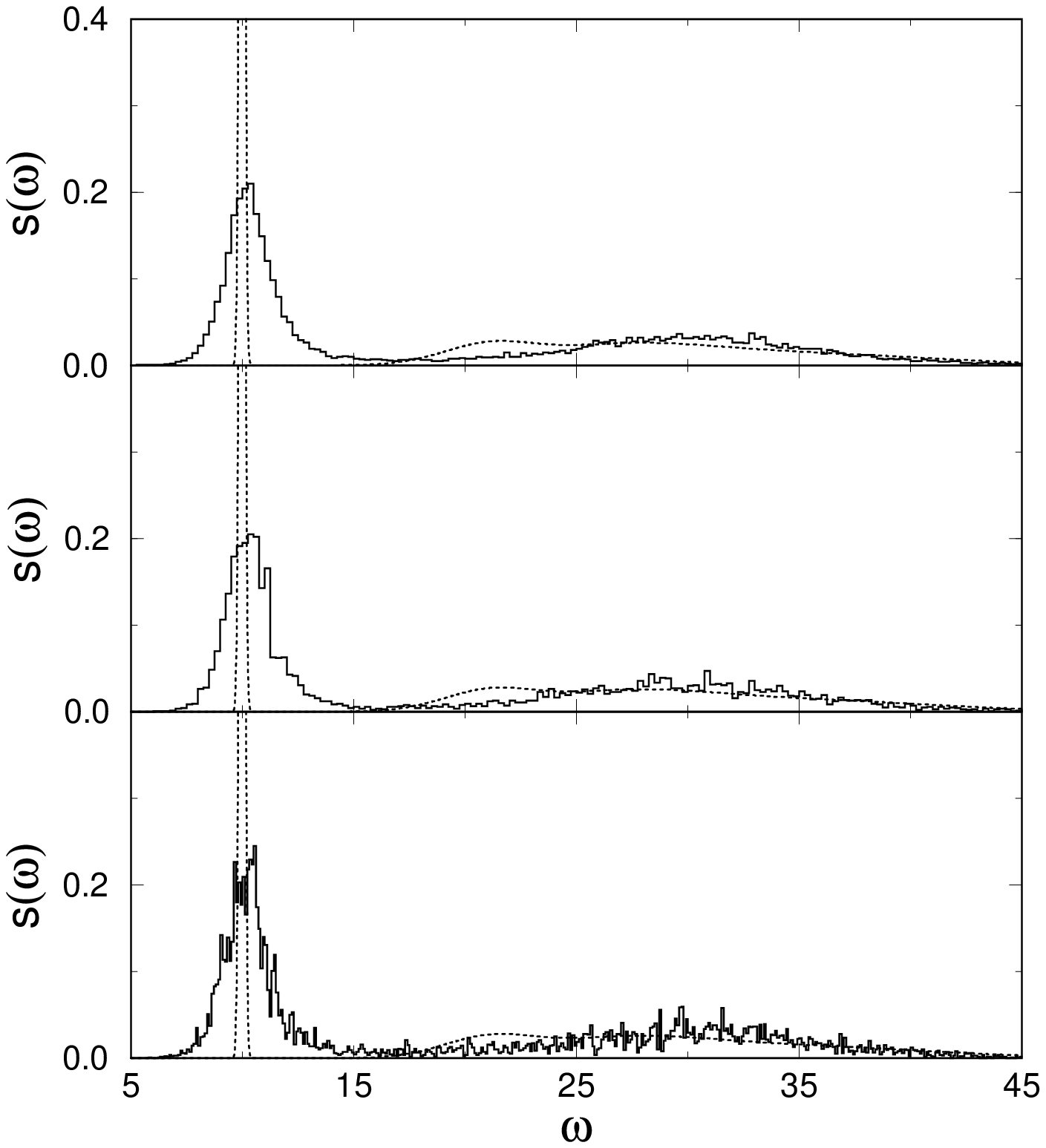}
 \caption{\label{graf6}
$s(\omega)$ (dotted line) as in Fig. \ref{graf4}.
Upper panel: reconstructed $S_{GIFT}(\omega)$ from the exact $f(\tau)$ (i.e. without noise)
assumed to be known for $l=60$ number of points in imaginary time with $\delta\tau=1/160$,
with $\Delta\omega=0.25$ and
using in the fitness $\Phi_{\mathcal{D}^{\star}}$ only the first moment (i.e. $\gamma_n=0$
$\forall n\neq 1$).
Middle panel: reconstructed $S_{GIFT}(\omega)$ from the exact $f(\tau)$ (i.e. without noise)
assumed to be known for $l=240$ number of points in imaginary time with $\delta\tau=1/640$,
with $\Delta\omega=0.25$ and
using in the fitness $\Phi_{\mathcal{D}^{\star}}$ only the first moment (i.e. $\gamma_n=0$
$\forall n\neq 1$).
Lower panel: reconstructed $S_{GIFT}(\omega)$ from the exact $f(\tau)$ (i.e. without noise)
assumed to be known for $l=60$ number of points in imaginary time with $\delta\tau=1/160$,
with $\Delta\omega=0.1$ and
using in the fitness $\Phi_{\mathcal{D}^{\star}}$ only the first moment (i.e. $\gamma_n=0$
$\forall n\neq 1$).
}
\end{figure}
Finally, we devise the following test: we try to reconstruct a spectral function
$s(\omega)$
($p_1=0.5$,      $p_2=0.1$,      $p_3=0.2$,      $p_4=0.2$,
 $\mu_1=10$,     $\mu_2=21$,     $\mu_3=27$,     $\mu_4=35$,
 $\alpha_1=0.1$, $\alpha_2=2$, $\alpha_3=4$, $\alpha_4=6$),
displaying a main peak and a broad contribution at higher $\omega$, made of
a superposition of three Gaussians, resembling qualitatively the {\it shape} of
the multiphononic contribution in the dynamical structure factor of superfluid $^4$He.
We have tested our strategy using the usual fitness function $\Phi_{\mathcal{D}^{\star}}$
with only the first moment included (i.e. $\gamma_n=0$ $\forall n\neq 1$):
the results are plotted in Fig. \ref{graf4}.
In Fig. \ref{graf4b} the integrated spectral functions are plotted;
from the comparison between the exact $I_0(\omega)$ and the one obtained from the
reconstructed $S_{GIFT}(\omega)$ one can observe that the spectral weights under
the main peak and the broad contribution are well reproduced.
Also the large $\omega$ limit of $I_{-1}(\omega)$ is in good agreement with the exact value.

One can also study the effect of the noise in $f(\tau)$
in order to check the GIFT ability in recovering correct information on 
the true $s(\omega)$.
In Fig. \ref{graf5} we show two $S_{GIFT}(\omega)$ reconstructed from a noisy $f(\tau)$
with $\sigma_{\varepsilon_j}$ 10 times and 50 times greater than in the test shown in Fig. \ref{graf4}.
Only in the second case, which represents a situation of very high relative noise
($\sigma_{\varepsilon_j}/f_j$ in the range 5--200 \%),
information on the correct spectral function is sensibly lost.
This test show the robustness of GIFT against noise in the observations, being able to
recover correct information on $s(\omega)$ with a noise level up to one order of magnitude
greater than what can be easily obtained in typical QMC calculations of imaginary--time
correlation functions.

It is possible also to use GIFT with a limited information on $f(\tau)$,
which corresponds as usual to $f(\tau)$ values for a discrete set of imaginary times,
but without any added noise.
In this case the average procedure in \eqref{reconstructed} consists of an average among
models found compatible with one single set $\mathcal{F}$.
The result of such GIFT multi--peak reconstruction is shown in the upper panel of Fig. \ref{graf6}.
By comparing this result with that shown in Fig. \ref{graf4} it is possible to see that the two
$S_{GIFT}(\omega)$ are very similar thus ruling out the necessity of more accurate observations
of $f(\tau)$ at discrete imaginary times in order to improve the GIFT performance.
By maintaining the noise level in $f_j$ to zero, we have also tried to increase the amount
of information by using $l=240$ number of points in imaginary time with $\delta\tau=1/640$;
the result of such GIFT multi--peak reconstruction is shown in the middle panel of Fig. \ref{graf6}.
No substantial improvement can be observed with respect to the previous case in spite of an increased
computational cost.
The computational cost of GIFT is increased also by considering a wider space of model spectral
functions. In our last test we tried a GIFT multi--peak reconstruction without noise
with $\Delta\omega=0.1$, the number of bins in frequency space  $N_{\omega}=1500$
and the ``quantization'' of spectral weight  $\mathcal{M}=10^4$.
The result is shown in the lower panel of Fig. \ref{graf6}; here the noise in $S_{GIFT}(\omega)$
is higher because due to the computational cost of the GIFT strategy with this parameters
we have only averaged over $\mathcal{N}_{r}=160$ random sets.
Also in this case we found no substantial improvement in $S_{GIFT}(\omega)$ as compared to the
standard case.


\begin{thebibliography}{99}
\bibitem{footnote} $\mathcal{D}^{\star}$ is obtained as before
by sampling independent Gaussian distributions centered on the original observations $\mathcal{D}$,
with variances which correspond to the estimated statistical uncertainties.
\end{thebibliography}
\end{document}